# Efficient, indistinguishable telecom C-band photons using a tapered nanobeam

Mohammad Habibur Rahaman[1,2,*], Samuel Harper[1,2,*], Chang-Min Lee[1,2,*], Kyu-Young Kim[3], Mustafa Atabey Buyukkaya[1,2], Victor J. Patel[4], Samuel D. Hawkins[4], Je-Hyung Kim[3], Sadhvikas Addamane[5], Edo Waks[1,2,6,7]

[1]Department of Electrical and Computer Engineering, University of Maryland, College Park, Maryland, 20742, USA.
[2]Institute for Research in Electronics and Applied Physics (IREAP), University of Maryland, College Park, Maryland, 20742, USA.
[3]Department of Physics, Ulsan National Institute of Science and Technology (UNIST), Ulsan, 44919, Republic of Korea.
[4]Sandia National Laboratories, Albuquerque, New Mexico, 87123, USA.
[5]Center for Integrated Nanotechnologies, Sandia National Laboratories, Albuquerque, NM 87123, USA
[6]Department of Physics, University of Maryland College Park, MD, 20742, USA.
[7]Joint Quantum Institute (JQI), University of Maryland, College Park, MD, 20742, USA.

**Corresponding author**, edowaks@umd.edu

**Abstract**
Telecom C-band single photons exhibit the lowest attenuation in optical fibers, enabling long-haul quantum-secured communication. However, efficient coupling with optical fibers is crucial for these single photons to be effective carriers in long-distance transmission. In this work, we demonstrate an efficient fiber-coupled single photon source at the telecom C-band using InAs/InP quantum dots coupled to a tapered nanobeam. The tapered nanobeam structure facilitates directional emission that is mode-matched to a lensed fiber, resulting in a collection efficiency of up to 65% from the nanobeam to a single-mode fiber. Using this approach, we demonstrate single photon count rates of 575 ± 5 Kcps and a single photon purity of $g^{(2)}(0) = 0.015 \pm 0.003$. Additionally, we demonstrate Hong-Ou Mandel interference from the emitted photons with a visibility of 0.84 ± 0.06. From these measurements, we determine a photon coherence time of 450 ± 20 ps, a factor of just 8.3 away from the lifetime limit. This work represents an important step towards the development of telecom C-band single-photon sources emitting bright, pure, and indistinguishable photons, which are necessary to realize fiber-based long-distance quantum networks.

**Introduction**
A telecom single-photon source operating around 1,550 nm (C-band) is essential for fiber-based quantum communication networks to minimize transmission loss in the silica fiber [1]. Various approaches have been explored to realize single-photon sources within this spectral window, including spontaneous parametric down-conversion [2], as well as materials such as carbon nanotubes [3] and semiconductor quantum dots [4,5]. Among these methods, single photon sources based on semiconductor quantum dots are considered most promising due to their high brightness [6,7], high purity [6,8], and indistinguishability [9-11]. For example, InAs/InP quantum dots emit single photons in the telecom C-band spectrum [8,10-17], which have been extracted utilizing various devices, such as mesa structures [11], planar cavities [16], nanowires [18], and photonic crystal cavities [7]. However, more efficient coupling to the optical fiber is needed to collect the telecom C-band single photon emission for long-distance quantum networks.

In this work, we demonstrate an efficient fiber-coupled single-photon source operating at the telecom C-band using InAs/InP quantum dots integrated in a tapered nanobeam waveguide. The nanobeam efficiently mode-matches the emission of the quantum dots into a lensed fiber, enabling the collection of the emitted light into a single-mode fiber with an efficiency of up to 65%. Using this device, we report bright single photon emission with a count rate exceeding $575 \pm 5$ Kcps. From second-order correlation measurements, we determine the photon purity to be $g^{(2)}(0) = 0.015 \pm 0.003$ at 350 nW. We also demonstrate the indistinguishability of the photon emission using Hong-Ou-Mandel two-photon interference measurements, with a visibility of $0.84 \pm 0.06$ and a photon coherence time of $450 \pm 20$ ps. These results pave the way for efficient fiber collection of single photons at optimal C-band wavelengths for long-distance quantum networks using a tapered nanobeam.

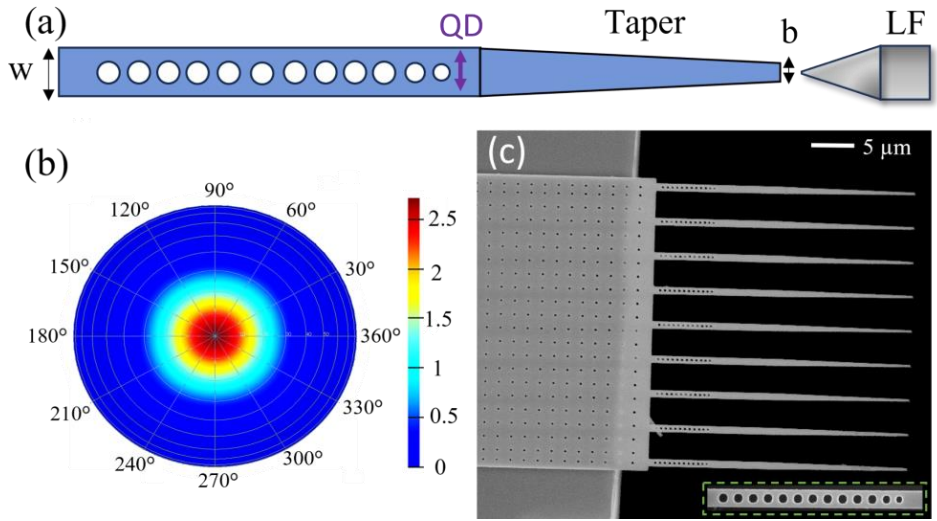

Figure 1: (a) Schematic diagram of the InAs/InP photonic crystal mirror nanobeam with width $w$, length $L$, and taper tip width of $b$. Here, QD: quantum dots; LF: lensed fiber. (b) The far-field mode profile was calculated at 500 nm from the tapered nanobeam tip edge. (c) SEM image of a fabricated nanobeam array after PDMS stamp transfer printing. The inset shows a zoomed-in image of a nanobeam mirror.

**Nanobeam Design and Fabrication**

The wafer was grown on a (100)-oriented Fe-doped InP substrate using molecular beam epitaxy [19,20]. The wafer consisted of a 280 nm thick InP membrane featuring InAs quantum dots at the center, grown on top of a 2 μm thick AlInAs sacrificial layer. The quantum dots were formed by first growing a 1.1 monolayer of InAs at a growth temperature of 510 °C. This layer was then annealed at the growth temperature for 120 seconds to facilitate the ripening technique [7,19], enabling control over the quantum dots density and emission wavelength at ~1550 nm. This process resulted in a quantum dot layer with densities between 10 and 50 μm$^{-2}$. We capped the quantum dots using a double-layer capping configuration, as described in [21].

To efficiently collect emission from the quantum dots, we employed a tapered nanobeam structure [22,23] featuring a periodic array of air holes, which serve as a Bragg mirror to ensure the quantum dot emission is directed towards the tapered end (Figure 1a) (details in Supplementary Section 1). The width of the nanobeam was adiabatically reduced from 560 nm to 150 nm over a 15 μm long taper to efficiently mode-match to a lensed fiber with a numerical aperture of 0.50. Figure 1b displays the 3D-FDTD simulated far-field mode profile of the tapered nanobeam, illustrating the directional emission from the edge of the adiabatic linear taper. Furthermore, using 3D-FDTD, we calculated an 88% coupling efficiency into the lensed fiber, based on the overlap of the far-field emission of the nanobeam taper with the Gaussian function of the fiber mode (details in Supplementary Section 1) [24].

To fabricate the designed nanobeam waveguide, we used a combination of electron beam lithography and transfer print lithography using previously reported techniques for nanobeams [22,25] and nanobeam cavities [26]. Figure 1c shows the transferred structure with the nanobeam array fully suspended from the edge of a cleaved silicon chip. The inset provides a closer view of the photonic crystal nanobeam mirror.

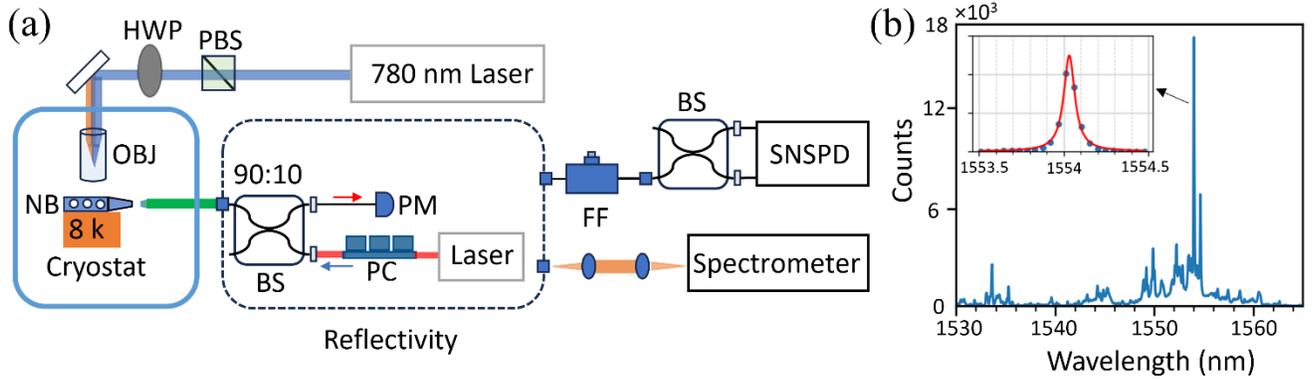

Figure 2: (a) Schematic diagram of the optical measurement setup. LF: lensed fiber, FF: fiber filter, NB: nanobeam, HWP: half waveplate, BS: beam splitter, PBS: polarizing beam splitter, OBJ: objective lens, PM: power meter, PC: polarization controller, and SNSPD: superconducting nanowire single photon detector. (b) Time-resolved photoluminescent spectra of quantum dot emission for an above-band excitation of 780 nm at a power 5 µW. The inset shows the zoomed-in spectrum of the QD emission filtered by a 0.2 nm bandwidth tunable fiber spectral filter. Here, the blue circles show the data, and the red solid line represents the Lorentzian fit.

**Results:**

Figure 2a shows the experimental setup we used to characterize the quantum dot emission. The sample was placed in a closed-cycle cryostat and cooled to 8 K. We aligned the tapered nanobeam to the lensed fiber using a low-temperature 3-axis nanopositioner array within the vacuum chamber. We excited the nanobeam using an above-band laser at an excitation wavelength of 780 nm, focused through an objective lens. After collecting from the lensed fiber, we filtered the light using a 0.2 nm bandwidth tunable fiber spectral filter for single photon counting and measurements. The filtered light was detected using a superconducting nanowire single-photon detector (SNSPD).

We first characterized the coupling efficiency of the nanobeam to the lensed fiber by measuring the direct reflectivity. We injected a tunable external cavity diode laser emitting at 1550 nm into the lensed fiber using a 90:10 fiber coupler and measured the amount of light back-reflected into the fiber with a power meter, as depicted in the dashed box in Figure 2a. Assuming that the input coupling and output coupling efficiencies are the same, we determined a 65% coupling efficiency into and out of the device (see details in Supplementary Section 2). We attribute the discrepancy between the simulated and observed efficiency to fabrication imperfections, angular misalignment between the fiber and nanobeam, and the fact that the photonic crystal mirror may introduce scattering losses.

Next, we characterized the quantum dot emission through photoluminescence. The quantum dots in the nanobeam were excited using the 780 nm laser in continuous wave mode at a power of 5 µW. We detected the emitted light using a spectrometer connected through the lensed fiber. Figure 2b shows the collected photoluminescence spectrum from a nanobeam in the telecom C-band window. The spectrum exhibits a series of sharp resonances, which we attribute to individual quantum dots. The inset provides a close-up view of an isolated quantum dot emission line at a wavelength of 1554.05 nm, filtered by a 0.2 nm bandwidth tunable fiber spectral filter. A Lorentzian fit of this emission indicates a line width of 0.067 nm, which is near the spectrometer resolution limit of 0.055 nm. We focused all subsequent experimental measurements on this specific quantum dot.

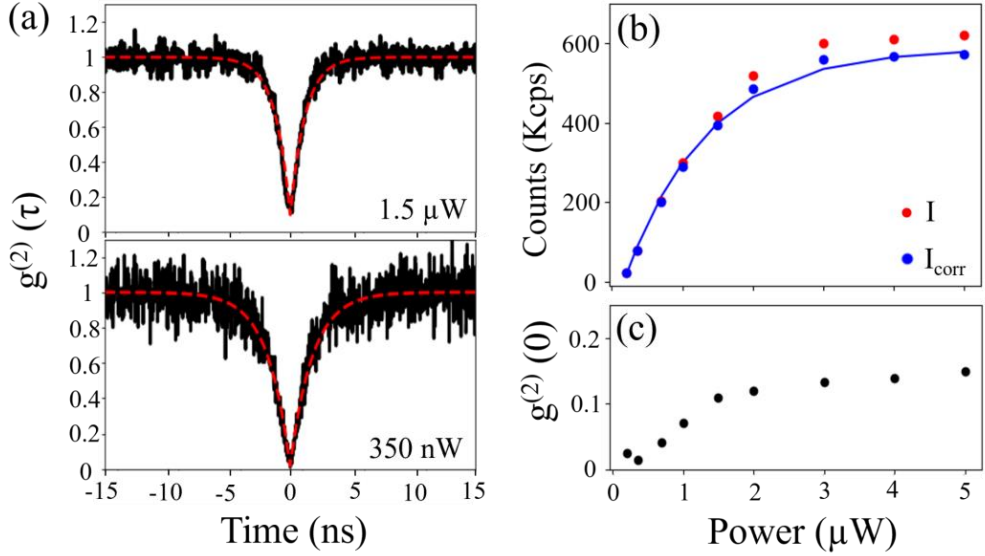

Figure 3: (a) Normalized autocorrelation histograms measured at 780 nm under above-band excitation with power levels of 350 nW and 1.5 µW. The black lines represent the measured data, while the red dashed lines depict curves fitted to the double-sided exponential decay function. (b) Detected count rates (red circles) and $g^{(2)}(0)$-corrected count rates (blue circles) as a function of the pump power. The blue line corresponds to a fit to the blue circles with fitting parameters $P_{sat} = 1.19$ µW and $I_{sat} = 575 \pm 5$ Kcps. (c) $g^{(2)}(0)$ as a function of the pump power.

To verify that the isolated emitter is a single photon source, we performed a second-order autocorrelation measurement. We excited the quantum dot within the nanobeam with the same continuous wave laser operating at 780 nm. In the collection path of the measurement setup, we used a fiber-type 50:50 beam splitter after the fiber filter and two superconducting nanowire single-photon detectors (Figure 2a). Figure 3a shows the measured second-order correlation for an excitation power of both 350 nW and 1.5 µW. The second-order autocorrelation shows clear anti-bunching, indicating that the emission corresponds to a single quantum dot. By numerically fitting the data to a double-sided exponential function, we determined $g^{(2)}(0) = 0.015 \pm 0.003$ at 350 nW and $g^{(2)}(0) = 0.109 \pm 0.004$ at 1.5 µW, which is exceptional for above-band excitation and could be further improved using quasi-resonant excitation [12]. From the fit to the second-order correlation, we also calculated the lifetime of the emitter as a function of the pump power-dependent antibunching decay rate constant [27,28] and extracted a natural radiative lifetime of $\tau_{rad} = 1.87 \pm 0.03$ ns (see details in Supplementary Section 3). This lifetime aligns with previously reported values for InAs/InP quantum dots at the telecom C-band [8].

To quantify the brightness of the quantum dot, we measured the emission intensity as a function of the pump power (Figure 3b). The red data points represent the detected count rates ($I$), which include both the quantum dot emission and background light sources, such as detector dark counts, background light, and broadband Raman emission from other quantum dots in the sample. It thus overestimates the brightness of the emitter. To properly extract the true count rate of the emitter, we also measured the single photon purity $g_2(0)$ for each pump power, as shown in Figure 3c. Using this value, we can calculate the count rate from the emitter only using the equation $I_{corr} = I\sqrt{1 - g^{(2)}(0)}$ [29], as shown by the blue data points in Figure 3b.

We numerically fit the corrected count rates to the saturation function $I(P) = I_0 + I_{sat}(1 - e^{-P/P_{sat}})$ [29], where $P$ is the pump power, $P_{sat}$ is the saturation power, $I_{sat}$ is the saturation intensity, and $I_0$ is the dark count, as shown by the blue curve in Figure 3b. At low powers, the emission rises linearly, but at high powers, the dot saturates and cannot produce a higher photon rate. From this fit, we determined the quantum dot saturation count rate to be 575 ± 5 Kcps (cps: counts per second). The count rates could be improved by incorporating the dot into a cavity and using the Purcell effect to radiatively enhance the emission [6,7].

To determine the source efficiency, we performed the same second-order autocorrelation measurement using a pulsed laser operating at 780 nm with a pulse duration of 300 ps and a repetition rate of 40 MHz (see Supplementary Section 4). We determined the saturation count rate under pulsed excitation to be $I_{sat}$ = 452 ± 4 Kcps, and by dividing this with the repetition rate of the pulsed laser, determined the total end-to-end efficiency to be 1.13 ± 0.01%. To assess the source efficiency, we performed a full photon budget of losses in the collection optics and detectors (see Supplementary Section 5). By accounting for these losses and using the end-to-end efficiency, we determined a source efficiency of 12.41 ± 0.09%. This efficiency is influenced by factors such as the non-radiative decay mechanisms of the quantum dot. Additionally, loss occurs as only the transverse electric (TE) polarizations of the emitter are efficiently coupled to the nanobeam waveguide mode [27]. This inefficiency can be mitigated by using direct resonant π-pulse excitation [30].

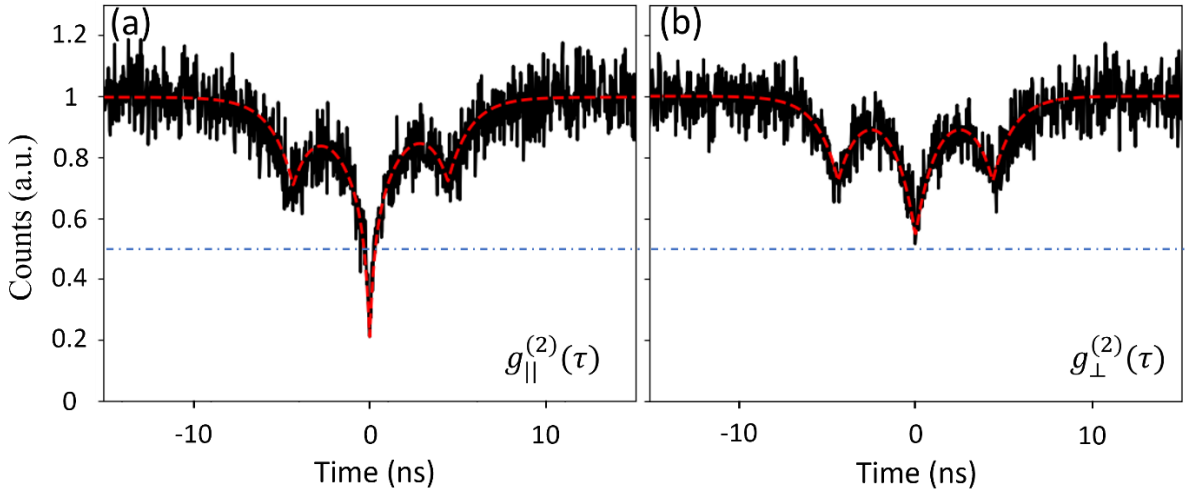

Figure 4: Two-photon interference for (a) indistinguishable photons with parallel polarization (co-polarized) and (b) distinguishable photons with orthogonal polarization (cross-polarized) at 1.5 µW pump power using a continuous wave laser at 780 nm, with respective fit functions represented by red dashed lines.

To determine the indistinguishability of the source, we performed a two-photon Hong-Ou-Mandel measurement using an unbalanced fiber-based Mach-Zender Interferometer in the collection path (see details in Supplementary Section 6). The quantum dot within the nanobeam was once again excited with 1.5 µW pump power, employing the same continuous-wave laser operating at 780 nm. The single photon generated experiences a time delay ($\Delta \tau_2$) of 4.36 ns at the first fiber-type beam splitter before eventually colliding at the second fiber-type beam splitter of the unbalanced Mach-Zender Interferometer, allowing us to determine indistinguishability.

Figures 4a and 4b show the measurement results of the Hong-Ou-Mandel experiment for co-polarized (indistinguishable) and cross-polarized (distinguishable) photons, respectively. In both cases, two dips near 0.75 at ± 4.36 ns are due to the delay introduced by the interferometer. The central dip corresponds to the two-photon interference at the second fiber-type beam splitter. In the case of indistinguishable photons, the central dip at zero-time delay falls below the classical limit of 0.5 [31], as indicated by the blue dashed horizontal line (Figure 4a), while distinguishable photons have a central dip near 0.5 (Figure 4b). The red dashed lines in Figure 4 represent correlation functions for the cross-polarized $g_\perp^{(2)}(\tau)$ and co-polarized $g_\parallel^{(2)}(\tau)$ cases for a continuous-wave Hong-Ou-Mandel experiment provided by [31,32].

$$g_\perp^{(2)}(\tau) = 4(T_1^2 + R_1^2)R_2 T_2 g^{(2)}(\tau) + 4 R_1 T_1 \left[T_2^2 g^{(2)}(\tau - \Delta \tau_2) + R_2^2 g^{(2)}(\tau + \Delta \tau_2)\right] \quad (2)$$

$$g_\parallel^{(2)}(\tau) = 4(T_1^2 + R_1^2)R_2 T_2 g^{(2)}(\tau) + 4 R_1 T_1 \left[T_2^2 g^{(2)}(\tau - \Delta \tau_2) + R_2^2 g^{(2)}(\tau + \Delta \tau_2)\right] \times$$
$$\left(1 - V e^{(-2|\tau|/\tau_c)}\right) \quad (3)$$

Where $R_{1,2}$ and $T_{1,2}$ denote the reflection and transmission probabilities of the two beam-splitters within the interferometer and $g^{(2)}(\tau)$ represents the second-order autocorrelation. In the co-polarized case, the additional terms include $\tau_c$, which stands for the coherence time of the single-photon source, and $V$, which represents the overlap of wave functions at the second beam splitter. From equation 3 for the co-polarized case, we determined the coherence time was $\tau_c = 450 \pm 20$ ps. This value is similar to the best reported results for C-band single photon sources under non-resonant and resonant excitation (440 ± 45 ps and 447 ± 15 ps) [16,17]. A high coherence time could enable high fidelity quantum information processing and increase the efficiency in repeater-based quantum communication systems at telecom C-band wavelengths [33].

We also evaluated the visibility of the indistinguishable photons based on coincidence counts for the co-polarized and cross-polarized photons. The visibility is defined as $V_{HOM} = 1 - \frac{g_\perp^{(2)}(0)}{g_\parallel^{(2)}(0)}$, derived from the zero-delay autocorrelation values of the co-polarized and cross-polarized cases. From Equations 2 and 3, the raw visibility was determined to be $V_{HOM} = 0.64 \pm 0.04$. However, visibility is constrained by the nonzero $g^{(2)}(0) = 0.109 \pm 0.004$ value at the 1.5 µW pump power, as shown in Figure 3a. Therefore, we corrected the visibility based on this $g^{(2)}(0)$ value, providing the upper bound of visibility as $M_S = \frac{V_{HOM} + g^{(2)}(0)}{1 - g^{(2)}(0)}$ [34,35]. Following the $g^{(2)}(0)$ correction, we found the visibility of indistinguishable photons was $M_S = 0.84 \pm 0.06$. This observed experimental visibility is lower than the ideal value ($M_S = 1$), primarily due to the finite time resolution of the detector, the timing jitter of our detector, and dark counts.

**Conclusion:**

In conclusion, we have demonstrated efficient fiber-coupled C-band indistinguishable photons using an InAs/InP quantum dot coupled to a tapered nanobeam. We achieved C-band single photons with a count rate of 575 ± 5 Kcps, $g^{(2)}(0) = 0.015 \pm 0.003$, two-photon interference visibility of 0.84 ± 0.06, and a photon coherence time of 450 ± 20 ps using above-band excitation. The efficiency of the device could be further improved by employing resonant cavities that increase radiative efficiency through the Purcell effect [7]. Moreover, utilizing resonant excitation [36], quasi-resonant excitation [37], and nanoscale focus pinspot [38] techniques could further reduce the $g^{(2)}(0)$ value and improve the indistinguishability by suppressing the time jitter induced by relaxation to the exciton ground state. In addition to reducing fiber losses, the current device structure can be coupled to integrated photonic circuits to serve as an on-chip single-photon source [39-42]. Ultimately, this tapered nanobeam design could enable new devices for quantum communication and distributed quantum computing by enabling the direct emission of telecom C-band photons without quantum frequency conversion [43].

**Author Contributions**

*M.H.R., *S. H. and *C-M. L. contributed equally to this work.

**Acknowledgment**
National Science Foundation (grants #OMA1936314, #OMA2120757), AFOSR grant #FA23862014072, the U.S. Department of Defense contract #H98230-19-D-003/008, Office of Naval Research (#N000142012551), and the Maryland-ARL Quantum Partnership (W911NF1920181). This work was performed, in part, at the Center for Integrated Nanotechnologies, an Office of Science User Facility operated for the U.S. Department of Energy (DOE) Office of Science. Sandia National Laboratories is a multimission laboratory managed and operated by National Technology & Engineering Solutions of Sandia, LLC, a wholly owned subsidiary of Honeywell International, Inc., for the U.S. DOE's National Nuclear Security Administration under contract DE-NA-0003525. The views expressed in the article do not necessarily represent the views of the U.S. DOE or the United States Government.

# Supplementary Information: Efficient, indistinguishable telecom C-band photons using a tapered nanobeam

Mohammad Habibur Rahaman[1,2,*], Samuel Harper[1,2,*], Chang-Min Lee[1,2,*], Kyu-Young Kim[3], Mustafa Atabey Buyukkaya[1,2], Victor J. Patel[4], Samuel D. Hawkins[4], Je-Hyung Kim[3], Sadhvikas Addamane[5], Edo Waks[1,2,6,7]

[1]Department of Electrical and Computer Engineering, University of Maryland, College Park, Maryland, 20742, USA.
[2]Institute for Research in Electronics and Applied Physics (IREAP), University of Maryland, College Park, Maryland, 20742, USA.
[3]Department of Physics, Ulsan National Institute of Science and Technology (UNIST), Ulsan, 44919, Republic of Korea.
[4]Sandia National Laboratories, Albuquerque, New Mexico, 87123, USA.
[5]Center for Integrated Nanotechnologies, Sandia National Laboratories, Albuquerque, NM 87123, USA
[6]Department of Physics, University of Maryland College Park, MD, 20742, USA.
[7]Joint Quantum Institute (JQI), University of Maryland, College Park, MD, 20742, USA.

**Description:** Supplementary Sections, Supplementary Figures, and Supplementary References.

## Section 1: Tapered Nanobeam Design

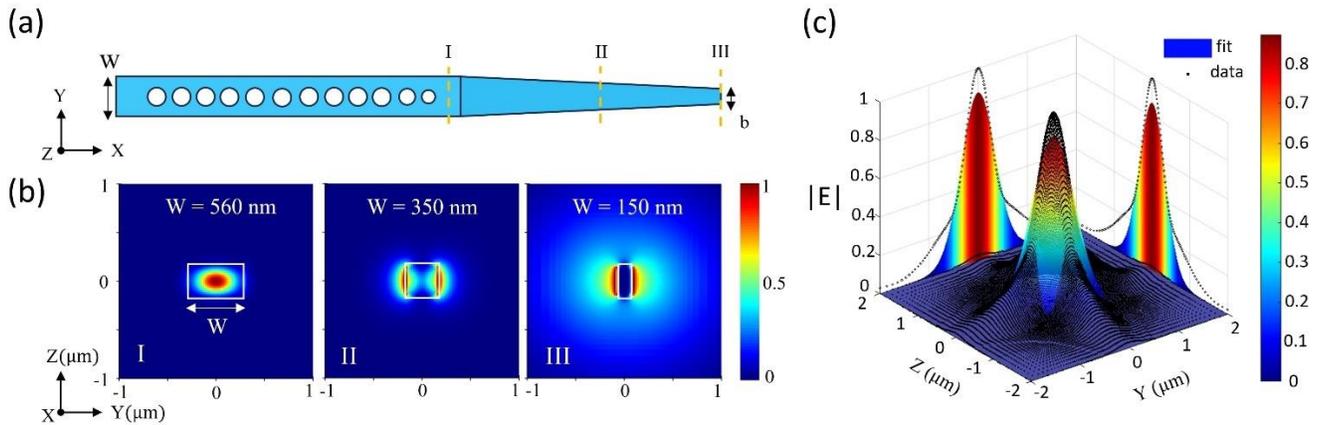

Figure S1: (a) A schematic diagram illustrating the design of the InAs/InP nanobeam with an adiabatic taper, where yellow dashed lines labeled as I-III indicate the positions of cross sections along the nanobeam taper. (b) A white rectangular box depicts the electric field intensity ($|E|$) profile for various cross sections (I-III) of the InAs/InP nanobeam taper. (c) The 3D far-field ($|E|$) profile at 500 nm away from the nanobeam taper edge. Here, the black dots represent the far-field profile of the nanobeam, and the color-coded surface map represents the Gaussian mode of the single-mode fiber.

Figure S1a illustrates the nanobeam design featuring an adiabatic taper with a thickness of 280 nm. Based on 3D finite-difference time-domain (FDTD) simulations, we designed the nanobeam with dimensions of 560 nm width and 12 air holes with a periodicity of 440 nm. The nanobeam features varying hole radii, with the first 10 holes at 110 nm and the last two at 95 nm and 80 nm, respectively. To design the nanobeam taper, we utilize 3D finite-difference time-domain (FDTD) simulations to analyze the electric field intensity profiles and far-field emission patterns of the nanobeam taper at 1550 nm, as shown in Figure S1b and c. In Figure S1b, the electric field intensity ($|E|$) is depicted for various cross sections labeled as I-III along the nanobeam taper. The taper width decreases linearly from 560 nm to 150 nm while maintaining a fixed nanobeam thickness of 280 nm over a length of 15 µm. As the taper width decreases, the waveguide mode gradually decays evanescently towards the outside of the waveguide, as observed at position III when the taper width is 150 nm.

The tapered structure expands the propagation mode and results in directional Gaussian-shaped emission from the tapered edge, as demonstrated in the 3D FDTD simulation represented by the black points in Figure S1c. The efficiency of collecting single photons depends on the overlap between the emission from the nanobeam taper and the Gaussian mode profile of the single-mode optical fiber. We utilized nonlinear least-squares fitting to approximate the nanobeam taper field profiles with Gaussian functions, as shown by the color-coded surface map of Figure S1c. To estimate the fiber coupling efficiency, we calculated the Gaussian overlap, using the following equation [22]

$$\eta_{fiber} = \frac{|\iint E_{nb} E_f^* \, dx \, dy|^2}{\iint |E_{nb}|^2 \, dx \, dy \iint |E_f|^2 \, dx \, dy}$$

where $E_{nb}$ and $E_f$ represent the field intensity of the nanobeam emission taper and fiber mode, respectively. Based on this equation, we calculate an 88% overlap between the field emission from the nanobeam taper and the fiber mode.

**Section 2: Reflectance Measurement**

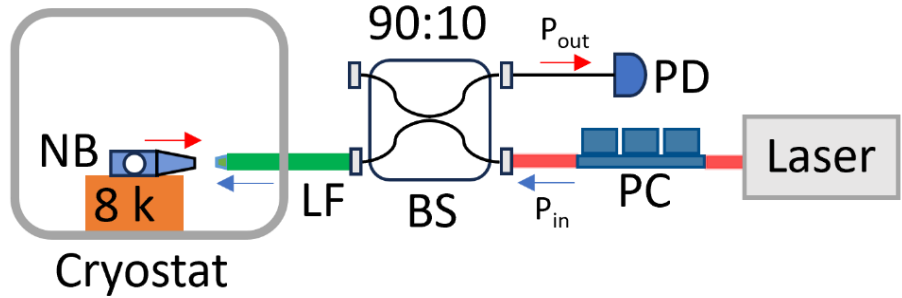

Figure S2. Schematic diagram of the reflectivity measurement setup. Here, NB: nanobeam, LF: lensed fiber, BS: beam splitter, PM: power meter, and PC: polarization controller.

Figure S2 illustrates the setup for measuring reflectivity to determine the coupling efficiency from the nanobeam to the lensed fiber. We assumed equal coupling-in and coupling-out efficiencies between the nanobeam and the lensed fiber. A laser with a power $P_{in}$ was injected into the lensed fiber, and the back-reflected light was measured as $P_{out}$ using a power meter. This measurement was performed with a 90:10 beam splitter having a transmission of $T = 83\%$. The coupling efficiency (η) is defined as

$$\eta = \sqrt{\frac{P_{out}}{T \times P_{in}}}.$$

## Section 3: Lifetime Calculation

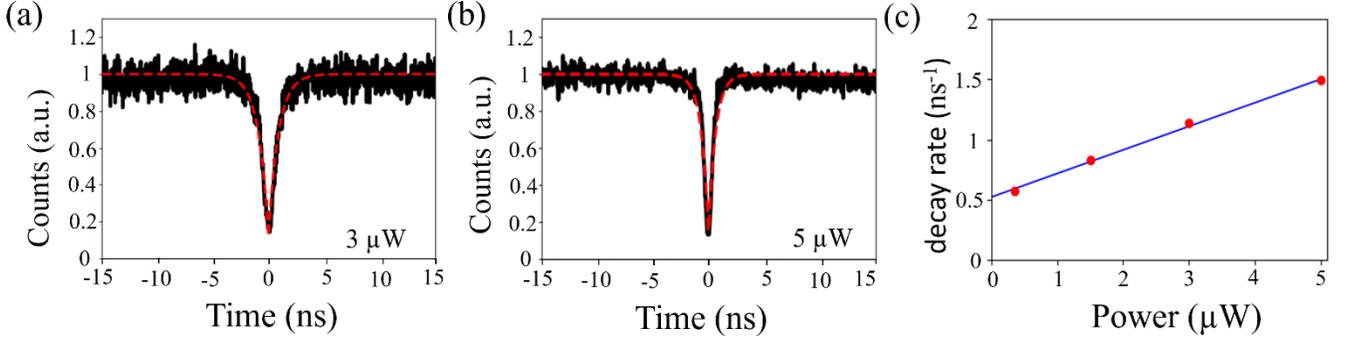

Figure S3: (a, b) Autocorrelation histograms measured at 780 nm under above-band excitation with power levels of (a) 3 µW and (b) 5 µW. The black lines represent the measured data, while the red dashed lines depict curves fitted to the double-sided exponential decay function. (c) The decay rate constant $\gamma_1$ at different pump powers obtained from the $g^{(2)}(\tau)$ histogram fits in (a) and (b) and Figure 3a of the main manuscript. The solid blue line is a fit to a linear function.

We determined the emitter's lifetime by calculating the decay rate of the antibunching dip observed in Figure 3a and Supplementary Figure S3a, b. Supplementary Figure S3a, b exhibit autocorrelation values of $g^{(2)}(0) = 0.133 \pm 0.004$ and $g^{(2)}(0) = 0.149 \pm 0.005$ at pump powers of 3 µW and 5 µW, respectively. The measured $g^{(2)}(\tau)$ was fitted to a two-level emitter second-order autocorrelation in the form of $g^{(2)}(\tau) = 1 - (1 - g^{(2)}(0))e^{-\gamma_1|\tau|}$. The decay rate constant $\gamma_1$ depends on the excited state lifetime $\tau_{rad}$ at pump power $P$, given by $\gamma_1 = 1/\tau_{rad}(1 + \alpha P)$, where $\tau_{rad}$ is the excited state lifetime, and $\alpha$ is a fitting parameter [27,28]. As the pump power increased, the decay rate constant also increased, as illustrated in Figure S3c. Fitting the data to a linear function, we obtained the excited state lifetime $\tau_{rad} = 1.87 \pm 0.03$ ns at zero power.

## Section 4: Second-order autocorrelation using pulsed laser excitation

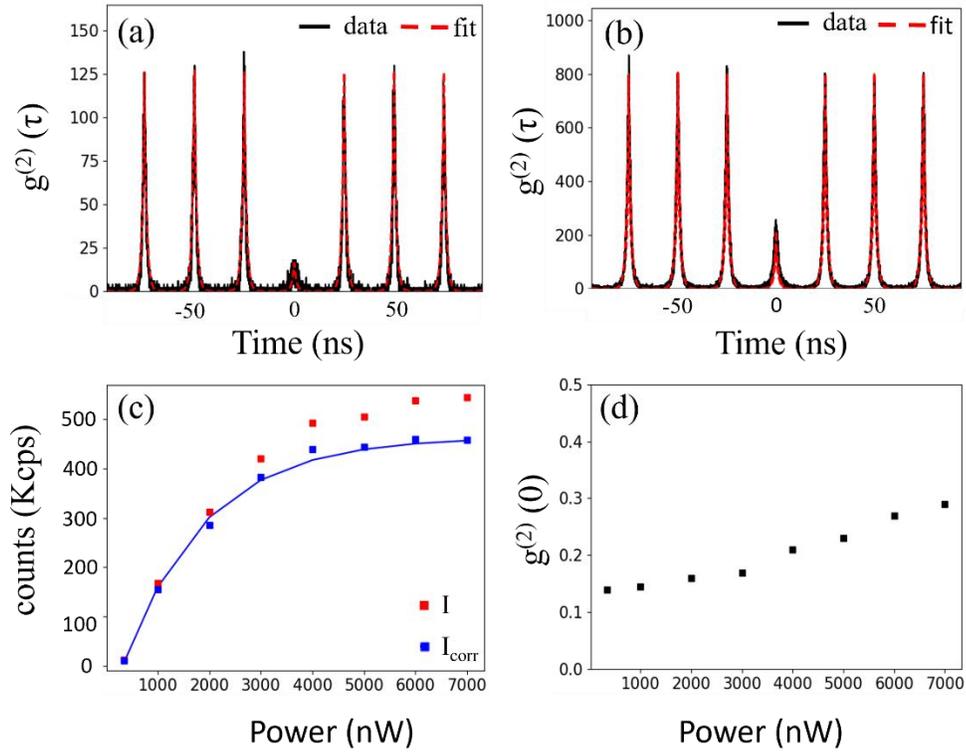

Figure S4: (a, b) Autocorrelation histogram measurements using at 780 nm laser under above-band pulsed laser excitation with power levels of (a) 250 nW and (b) 5 µW. The black lines represent the measured data, while the red dashed lines depict curves fitted to the double-sided exponential decay function. (c) Detected count rates (red squares) and $g^{(2)}(0)$-corrected count rates (blue squares) as a function of the pump power. The blue line corresponds to a fit with fitting parameters $P_{sat} = 1.6$ µW and $I_{sat} = 452 \pm 4$ Kcps. (d) $g^{(2)}(0)$ as a function of the pump power.

In addition to continuous wave excitation, we conducted second-order autocorrelation measurements for an above-band 780 nm pulsed laser with a 40 MHz repetition rate and a 300 ps pulse width. Autocorrelation histograms were measured at power levels of 250 nW and 5 µW, resulting in $g^{(2)}(0) = 0.135 \pm 0.01$ at 250 nW and $g^{(2)}(0) = 0.27 \pm 0.02$ at 5 µW, as illustrated in Figure S4a, b. Figure S4c depicts the count rate from the emitter as a function of the pump intensity. The raw counts vs. power and the $g^{(2)}(0)$-corrected fit $g^{(2)}(0)$(shown in Figure S4d) are presented with fitting parameters $P_{sat} = 1.6$ µW and $I_{sat} = 452 \pm 4$ Kcps.

## Section 5: Photon budget and source efficiency

We conducted a comprehensive photon budget analysis to assess the source efficiency. The total end-to-end efficiency of the system was 1.13 ± 0.01%, which was determined by dividing the count rate of the emitter at saturation ($I_{sat}$ = 452 ± 4 Kcps, Figure S4c) by the laser pulse repetition rate of 40 MHz. We also measured the efficiency of the individual system components and found (i) the lensed fiber coupling efficiency, $\eta_F$ = 0.6, (ii) the fiber-based spectral filter transmission efficiency, $\eta_{SF}$ = 0.526, (iii) the 50:50 beam splitter transmission efficiency, $\eta_{BS}$ = 0.44, (iv) the fiber cable transmission efficiency, $\eta_{FC}$ = 0.82, and (v) the SNSPD detection efficiency $\eta_{SNSPD}$ = 0.8. Utilizing these component transmission efficiencies, the efficiency of the lensed fiber was computed as $B_{fib} = \frac{I_{sat}}{\eta_F \eta_{SF}} = 3.58 \pm 0.03\%$, and the source efficiency was determined to be $B_{Source} = \frac{I_{sat}}{\eta_F \eta_{SF} \eta_{BS} \eta_{FC} \eta_{SNSPD}} = 12.41 \pm 0.09\%$.

## Section 6: Hong-Ou-Mandel interferometer setup

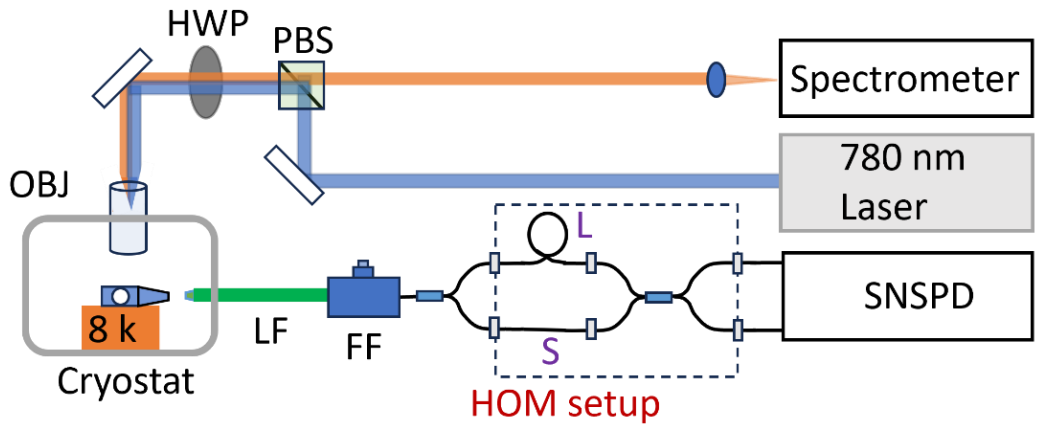

Figure S5: The Hong-Ou-Mandel (HOM) interferometer setup. The components include LF: lensed fiber, FF: fiber filter, HWP: half waveplate, PBS: polarizing beam splitter, OBJ: objective lens, and SNSPD: superconducting nanowire single-photon detector. L and S represent the long and short fiber paths introduced to create a time difference of 4.36 ns.

In the Hong-Ou-Mandel (HOM) two-photon interference measurements, the sample was excited using an above-band laser emitting light at a wavelength of 780 nm, focused with a zoom lens (OBJ), consistent with Figure 2a in the main manuscript. After collection from the lensed fiber, the light underwent initial filtration using a 0.2 nm grating tunable fiber filter designed for single photon counting. Fiber-type unbalanced Mach-Zehnder interferometers were inserted at collection paths within the dashed-line boxes shown in Figure S5. These interferometers featured a 4.36 ns time delay on one path relative to the other, effectively delaying generated single photons within the first fiber-type beam splitter before colliding at the second fiber-type beam splitter. For the Hong-Ou-Mandel measurement, a fiber-type polarization controller and polarization-maintained fibers were utilized to prepare single photons at both parallel and orthogonal polarizations, with the flexibility to interchange them independently. Single photon count rates and second-order autocorrelation measurements were conducted using a single optical fiber on the collection path. The same setup was used, excluding the dashed-line boxes in Figure S5a, to conduct the Hanbury Brown and Twiss-type measurements and calculate the photon count rate.